\documentclass[12pt,a4paper]{article}
\usepackage{epsfig}
\usepackage{amsmath}
\usepackage{amssymb}
\usepackage{subfigure}
\begin{document}
\title{The scalars from the topcolor scenario and the  spin
correlations of the top pair production at the LHC }
\author{Chong-Xing Yue, Ting-Ting Zhang, Jin-Yan Liu\\
%\thanks{E-mail:cxyue@lnnu.edu.cn} \\
{\small Department of Physics, Liaoning  Normal University, Dalian
116029, P. R. China}
\thanks{E-mail:cxyue@lnnu.edu.cn}}
\date{\today}

\maketitle
\begin{abstract}

\vspace{1cm}

The topcolor scenario predicts the existences of some new scalars.
In this paper, we consider the contributions of these new particles
to the observables, which are related to the top quark pair
($t\overline{t}$) production at the LHC. It is found that these new
particles can generate significant corrections to the $t\bar{t}$
production cross section and the $t\bar{t}$ spin correlations.

 \vspace{2.0cm} \noindent

\end{abstract}
\newpage
\noindent{\bf 1. Introduction} \vspace{0.5cm}

Searching for the standard model (SM) Higgs boson is one of the main
tasks of the large hadron collider (LHC), which has a considerable
capability to discover and measure almost all of its quantum
properties [1]. However, if the LHC finds evidence for a new scalar
state, it may not necessarily be the SM Higgs boson. Most of the new
physics models beyond the SM predict the existence of new scalar
states. These new particles may produce contributions to some
physical observables. Thus, studying the possible signals of the new
scalar states at the current and near future high energy collider
experiments is of special interest, which will be helpful to test
the SM and further to differentiate various kinds of new physics
models.

To completely avoid the problems of triviality and hierarchy arising
from the elementary Higgs field in the SM, various kinds of
dynamical electroweak symmetry breaking (EWSB) models have been
proposed. Among these new physics models, topcolor scenario is
attractive because it can explain the large top quark mass and
provides a new possible EWSB mechanism [2]. The main features of
this kind of model are: EWSB is mainly driven by the technicolor
interaction, the masses of the light quarks and leptons and a very
small portion of the top quark mass are generated by the extended
technicolor interaction. The topcolor interaction gives the main
part of the top quark mass and makes small contributions to EWSB.
Topcolor scenario generally predicts a number of new scalar states
at the electrowake scale: three top-pions ($\pi_t^\pm, \pi_t^0$), a
top-Higgs boson($h_t^{0}$), and a techni-Higgs  boson ($h^0_{tc}$).
Some of these new particles couple preferentially to the third
generation fermions and might produce significant contributions to
the physical observables related to the top quark.

The top quark with a mass of the order of the electroweak scale is
the heaviest elementary particle discovered to date, which is
singled out to play a key role in probing the new physics beyond the
SM [3]. An important property of the top quark is that, compared to
lighter quarks, its lifetime is extremely short so that its
properties are not polluted by the hadronization process. In the
absence of hard gluon radiation, top quark polarization is
conserved, its spin information can be transferred  into its decay
products. This information can be used to study the Lorentz
structure of interaction vertices involved in top quark production
and decay.

At the LHC, a large number of top quarks will be produced every
year. This fact makes that it is possible to measure the observables
that depend on the top quark spin and the top quark properties with
high precision at this facility, which will provide a good probe for
tests of the SM and for searches of new physics beyond the SM. For
hadronic top quark pair ($t\bar{t}$) production, spin correlations
have been extensively studied in the quantum choromodynamics (QCD)
[4, 5, 6]. If the new particles have sizable couplings to the top
quark, then they can produce contributions to the $t\bar{t}$ spin
correlations. Effects of new physics on the $t\bar{t}$ spin
correlations have been studied at the $e^+e^-$ collider [7], the
photon collider [8], and the hadron colliders [9]. In this paper, we
consider the contributions of the scalar particles predicted by the
topcolor scenario to the $t\bar{t}$ production at the LHC and
further discuss  their effects on the $t\bar{t}$ spin correlations.
We find that these new particles can indeed give significant
contributions to the $t\bar{t}$ production cross section and sizable
deviation of the $t\bar{t}$ spin correlations from the SM prediction
are possible with reasonable values of the free parameters.

The topcolor-assisted technicolor (TC2) model [10] is one of the
phenomenologically viable models, which has almost all essential
features of the topcolor scenario. So, in the rest of this paper, we
will give our numerical results in detail under this model. In the
following section, we will give the relevant formula for our
calculation. The numerical results and a short discussion are shown
in the last section.

\vspace{0.5cm} \noindent{\bf 2. The relevant calculation formula }
\vspace{0.5cm}

For the TC2 model [10], technicolor (TC) interactions play a main
role in breaking the electroweak symmetry. Topcolor interaction
makes small contributions to $EWSB$, and gives rise to the main part
of the top quark mass, $(1-\varepsilon)m_{t}$, with the parameter
$\varepsilon\ll1$. Thus, there is the relation
\begin{equation}
\nu^{2}_{\pi}+F^{2}_{t}=\nu^{2}_{w},
\end{equation}
\vspace*{0.8cm}
\begin{figure}[htb]
\begin{center}
\vspace*{-0.4cm} \epsfig{file=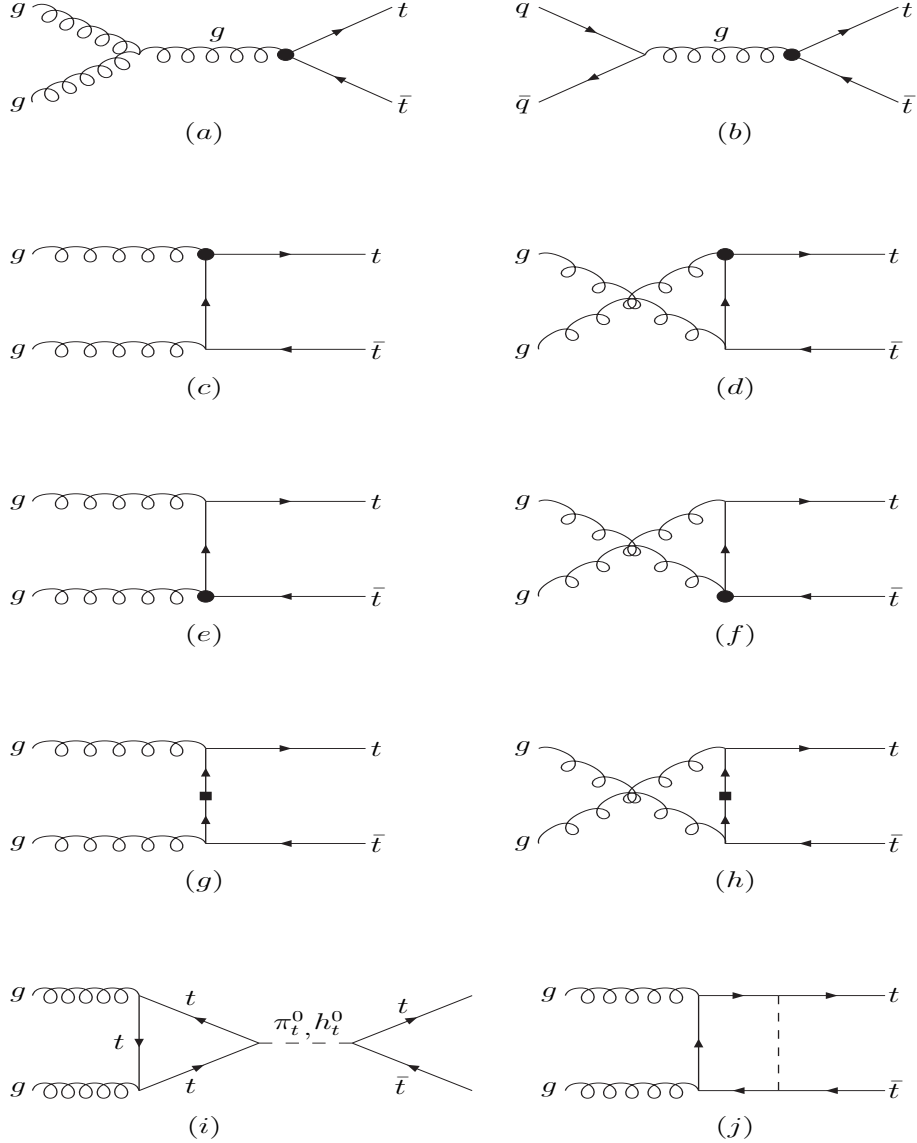,width=340pt,height=431pt}
\caption{Feynman diagrams for the partonic processes
$i$$\rightarrow$ $t\bar{t}$ ($i=gg$ and $q\bar{q}$) in the TC2
\hspace*{1.8cm}model.}
\end{center}
\end{figure}
where $\nu_{\pi}$ represents the contributions of TC interactions to
EWSB, $\nu_{w}=\nu/\sqrt{2}\simeq 174GeV$. Here $F_{t}\simeq 50GeV$
is the physical top-pion decay constant, which can be estimated from
the Pagels-Stokar formula. The scalar particles predicted by the TC2
model are bound-states of the techni-fermions and of top quark,
bottom quark, which are the techni-Higgs  boson ($h^0_{tc}$),
top-pions ($\pi_t^\pm, \pi_t^0$), and top-Higgs boson ($h^0_t$).

As it is well known that, in the TC2 model, topcolor interaction is
not flavor-universal and mainly couple to the third generation
fermions. Thus, the couplings of the top-pions ($\pi_t^\pm,
\pi_t^0$) to the three family fermions are non-universal and they
have large Yukawa couplings to the third family. The couplings of
these new scalar particles to the third family quarks can be written
as [10, 11]:
\begin{eqnarray}
\frac{m_t}{\sqrt{2}F_t}\frac{\sqrt{\nu_w^2-F_t^2}}{\nu_w}(iK^{tt*}_{UL}K^{tt}_{UR}
\bar{t}_Lt_R\pi^0_t+\sqrt{2}K^{tt*}_{UR}K^{bb}_{DL}\bar{t}_R b_L\pi^+_t+ \nonumber \\
iK^{tt*}_{UL}K^{tc}_{UR}
\bar{t}_L c_R\pi^0_t+\sqrt{2}K^{tc*}_{UR}K^{bb}_{DL}\bar{c}_R b_L\pi^+_t+h.c.),
\end{eqnarray}
where the factor $\sqrt{\nu_w^2-F^2_t}/\nu_w$ reflects the effect of
the mixing between the top-pions and the would be Goldstone bosons
[12]. To yield a realistic form of the $CKM$ matrix $V_{CKM}$, it
has been shown that the values of the matrix elements
$K_{UL(R)}^{ij}$ can be taken as [13]:
\begin{eqnarray}
K^{tt}_{UL}\approx K^{bb}_{DL}\approx1,\hspace{0.5cm}
K^{tt}_{UR}\approx1-\varepsilon, \hspace{0.5cm}K_{UR}^{tc}\leq\sqrt{2\varepsilon-\varepsilon^{2}}.
\end{eqnarray}

The relevant couplings for the top-Higgs boson ($h^0_t$) are similar
to those of the neutral top-pion $\pi^0_t$ [13]. However, for the
techni-Higgs  boson $h^0_{tc}$, it is not this case. Its coupling to
the top quark pair $t\bar{t}$ is very small, which is proportionate
to the factor $ \varepsilon/\sqrt{2}$ [14]. Furthermore, the mass of
$h^0_{tc}$ is at the order of $1TeV$. Thus, compared to the
top-Higgs and the top-pions, the contributions of the techni-Higgs
to the $t\bar{t}$ production can be neglected.

At hadron colliders, the top quark pair $t\bar{t}$ is produced
through the partonic processes of quark-antiquark pair annihilation
and gluon fusion:
\begin{eqnarray}
i\rightarrow t+\bar{t},\hspace{0.8cm} i=gg , \hspace*{0.6cm}q\bar{q}.
\end{eqnarray}
Where $q$ denotes $u$, $c$, $d$, $s$, or $b$ quark. It is well known
that the former is the dominate process at the LHC. In the context
of the TC2 model, the Feynman diagrams for these partonic processes
 are depicted in Fig.1. The black dot in Fig.1 represents the effective $gt\bar{t}$ vertex induced by the new
scalar particles ($\pi^\pm_t$, $\pi^0_t$, $h^0_t$), as shown in
Fig.2. Note that the boson in each loop denotes a neutral top-pion,
top-Higgs or a charged top-pion, while the fermion in each loop can
be a top or bottom quark depending on the involved boson being
neutral or charged. \vspace*{0.8cm}
\begin{figure}[htb]
\begin{center}
\vspace{-0.6cm} \epsfig{file=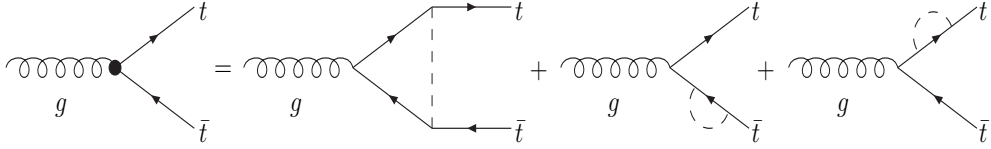,width=370pt,height=55pt}
\caption{Feynman diagrams for the effective vertex
 $gt\bar{t}$ in the TC2 model.}
\end{center}
\end{figure}

The invariant amplitudes for the partonic processes $g(p_{1})
g(p_{2})\rightarrow t(k_{1}, s_{t})\bar{t}(k_{2}, s_{\bar{t}})$ and
$q(p '_{1}) \bar{q}(p '_{2})\rightarrow t(k_{1},
s_{t})\bar{t}(k_{2}, s_{\bar{t}})$ contributed by the neutral
top-pion or the top-Higgs can be written as follows:
\begin{eqnarray}
M_{a}&=&g_{s}^{2}T^{a}f^{abc}(p_{1}-p_{2})_{\sigma}\epsilon^{\mu}(p_{1})
\epsilon^{\nu}(p_{2})\frac{g_{\mu\nu}}{(p_{1}+p_{2})^{2}}\bar{u}
(k_{1},s_{t})[A_{a}\gamma^{\sigma}
+iB_{a}\sigma^{\sigma\alpha}(p_{1}+p_{2})_{\alpha}\nonumber\\&&+C_{a}(\gamma^{\sigma}
-\frac{2m_{t}}{(p_{1}+p_{2})^{2}}(p_{1}+p_{2})^{\sigma})]\nu(k_{2},s_{\bar{t}});
\end{eqnarray}
\begin{eqnarray}
M_{b}&=&-ig_{s}^{2}T^{a}T^{b}\bar{v}(p'_{1})\gamma^{\mu}u(p'_{2})\frac{g_{\mu\nu}}{(p'_{1}+p'_{2})^{2}}
\bar{u}(k_{1},s_{t})[A_{b}\gamma^{\nu}+iB_{b}\sigma^{\nu\alpha}(p'_{1}+p'_{2})_{\alpha}
\nonumber\\&&+C_{b}(\gamma^{\sigma}-\frac{2m_{t}}{(p'_{1}+p'_{2})}(p'_{1}+p'_{2})^{\sigma})]\nu(k_{2},s_{\bar{t}});
\\\nonumber\\
M_{c}&=&-ig_{s}^{2}T^{a}T^{b}\epsilon^{\mu}(p_{1})
\epsilon^{\nu}(p_{2})\bar{u}(k_{1},s_{t})[A_{c}\gamma_{\mu}+B_{c}(k_{1}-p_{1})_{\mu}\not\!\!p_{1}
+C_{c}(k_{1}-p_{1})_{\mu}\gamma^{5}\nonumber\\
&&+D_{c}(k_{1}-p_{1})_{\mu}\not\!\!p_{1}\gamma^{5}+
E_{c}(k_{1}-p_{1})_{\mu}+ F_{c}\gamma_{\mu}\gamma^{5}
+G_{c}\not\!\!p_{1}\gamma_{\mu}+H_{c}\not\!\!p_{1}\gamma_{\mu}\gamma^{5}]\nonumber\\&&
\frac{(\not\!\!k_{1}-\not\!\!p_{1}+m_{t})}{(k_{1}-p_{1})^{2}-m_{t}^{2}}\gamma_{\nu}\nu(k_{2},s_{\bar{t}});
\end{eqnarray}
\begin{eqnarray}
M_{e}&=&-ig_{s}^{2}T^{a}T^{b}\epsilon^{\mu}(p_{1})
\epsilon^{\nu}(p_{2})\bar{u}(k_{1},s_{t})\gamma_{\mu}\frac{(\not\!\!k_{1}-\not\!\!p_{1}+m_{t})}{(k_{1}-p_{1})^{2}-m_{t}^{2}}
[A_{e}\gamma_{\nu}+B_{e}(k_{1}-p_{1})_{\nu}\not\!\!p_{2}
\nonumber\\&&+C_{e}(k_{1}-p_{1})_{\nu}\gamma^{5}+D_{e}
(k_{1}-p_{1})_{\nu}\not\!\!p_{2}\gamma^{5}+
E_{e}(k_{1}-p_{1})_{\nu}+ F_{e}\gamma_{\nu}\gamma^{5}
\nonumber\\&&+G_{e}\not\!\!p_{2}\gamma_{\mu}+H_{e}\not\!\!p_{2}\gamma_{\mu}\gamma^{5}]
\nu(k_{2},s_{\bar{t}});
\end{eqnarray}
\begin{eqnarray}
M_{g}&=&g_{s}^{2}T^{a}T^{b}\epsilon^{\mu}(p_{1})\epsilon^{\nu}(p_{2})
\bar{u}(k_{1},s_{t})\gamma_{\mu}\frac{\not\!\!k_{1}-\not\!\!p_{1}+m_{t}}{(k_{1}-p_{1})^{2}-m_{t}^{2}}
[\tilde{\Sigma}(\not\!\!k_{1}-\not\!\!p_{1})
+(\not\!\!k_{1}-\not\!\!p_{1})\nonumber\\
&&(\delta Z_{V}-\delta Z_{A}\gamma_{5}) +m_{t}\delta
m_{t}]\frac{\not\!\!k_{1}-\not\!\!p_{1}+m_{t}}{(k_{1}-p_{1})^{2}-m_{t}^{2}}\gamma_{\nu}\nu(k_{2},s_{\bar{t}});
\end{eqnarray}
\begin{eqnarray}
M_{i}&=&-g^2_s\epsilon^\mu(p_1)\epsilon^\nu(p_2)[A_ig_{\mu\nu}
+B_i(p_2-p_1)_\mu(p_2-p_1)_\nu+C_i(p_2-p_1)_\mu p_{1\nu}\nonumber\\
&&+D_i(p_2-p_1)_\mu(p_1+p_2)_\nu+E_i(p_1+p_2)_\mu(p_2-p_1)_\nu+F_i(p_1+p_2)_\mu
p_{1\nu}\nonumber\\
&&+G_i(p_1+p_2)_\mu(p_1+p_2)_\nu+H_i(p_1+p_2)^\mu(p_1+p_2)^\nu\varepsilon_{\mu\alpha\nu\sigma}
+I_i(p_1+p_2)_\mu p_{1\nu}\nonumber\\
&&+J_i(p_1+p_2)^\alpha
p_1^\sigma\varepsilon_{\mu\alpha\nu\sigma}]\frac{1}{(p_1+p_2)^2-m^2_t}\bar{u}(k_1,
s_t)(g_V+g_A\gamma_5)\nu(k_2, s_{\bar{t}}).
\end{eqnarray}
Here $g_s$ is the QCD coupling constant, $T^{i}$ stands for the
color generator. $k_1$($s_t$) and $k_2$($s_{\bar{t}}$) denote the
momentum(spin) of the top and anti-top quark, respectively.

For the renormalization of the ultraviolet divergences appearing in
the evaluation of the vertex and fermion self-energy corrections, we
 have used the on-mass-shell renormalization scheme. The wave
function renormalization constants can be determined from the top
quark self-energy diagrams, which can be written as:
\begin{eqnarray}
\tilde{\Sigma}(\not\!\!p)=\not\!\!p[-(g_{A}^{2}+g_{V}^{2})B_{1}(p^{2})+2g_{A}g_{V}B_{1}(p^{2})\gamma_{5}]
+}{m_{t}(g_{V}^{2}-g_{A}^{2})B_{0}(p^{2}),
\end{eqnarray}
with
\begin{eqnarray}
B_{1}=\frac{1}{2p^{2}}[A_{0}(m_{t}^{2})-A_{0}(M^{2})-(m_{t}^{2}-M^{2}+p^{2})B_{0}],
\hspace*{0.7cm}
B_{0}=B_{0}(p^{2},m_{t}^{2},M^{2}).
\end{eqnarray}
Here $M$ denotes the mass of the scalar particle  and $p$ is the
momentum of the top quark $t$.  $A_{0}$,
 and $B_{0}$  are the well-known one-point, two-point and three
point scalar functions [15], which are given in terms of the
Passarino-Veltman scalar functions.  The expression forms of the
form factors $A_i$ and $B_i$, etc have been given in the Appendix
$A1 \sim A5$. In the on-shell scheme, the finite parts of the
counter terms are determined by the requirement that the residue of
the fermion propagator is equal to one, which fixes the wave
function renormalization constraints $\delta Z_{V}$, $\delta Z_{A}$,
and $\delta m_{t}$. Their expression forms are given in Appendix A6.
The vector- and axis- vector coupling constants $g_V$ and $g_A$ can
be written as:
\\
for the neutral top-pion $\pi_t^0$, $g_V=0$, \hspace*{0.7cm}
$g_A=(1-\varepsilon)m_t\sqrt{\nu_w-F_t^2}/\sqrt{2}F_t\nu_w$;
\\
for the top-Higgs boson $h_t^0$,
$g_V=-i(1-\varepsilon)m_t\sqrt{\nu_w-F_t^2}/\sqrt{2}F_t\nu_w$,\hspace*{0.7cm}
$g_A=0$.

The invariant amplitudes $M_d$, $M_f$, $M_h$ are same as $M_c$,
$M_e$, $M_g$, respectively, but with $p_1\rightarrow p_2$. Because
the expression form of $\mathcal {M}_{j}$ is lengthy, we do not
present them here. Although, compared with the contributions for
other diagrams, the contributions of Fig.1(j) to the $t\bar{t}$
production are small, our numerical results will include its
contributions.

Similar as above, we can give the invariant amplitudes for the
partonic processes $gg\rightarrow t\bar{t}$ and $q\bar{q}\rightarrow
t\bar{t}$ contributed by the charged top-pions $\pi_t^\pm$. In order
not to make this paper too long, we do not present their explicit
expressions here. However, in our calculation, we will include the
contributions of the charged top-pions.

Using above amplitudes, it is straightforward to calculate the cross
section $\hat{\sigma}^i(\hat{s})$ for the partonic process
$i\rightarrow t\bar{t}$. The corresponding hadronic cross section
$\sigma^i(\hat{s})$ can be obtained by folding
$\hat{\sigma}^i(\hat{s})$ with the parton distribution functions
(PDFs). The differentical cross section for the total hadronic cross
section $\sigma(s)$ is given by:
\begin{equation}
\frac{d\sigma(s)}{dX}=\sum_i\int dx_{1}dx_{2}
f_{i/p}(x_{1},\mu_{F})f_{i/p}(x_{2},\mu_{F})
\frac{d\hat{\sigma}^i(\hat{s})}{dX},
\end{equation}
where $i=q\bar{q}$ or $gg$ and $X$ can be chosen to be the
$t\bar{t}$ invariant mass $M_{t\bar{t}}$ or the transverse momentum
$P_T$ of the top quark. $\hat{s}=x_{1}x_{2}s$ is the effective
center-of-mass (c.m.) energy squared for the partonic process
$i\rightarrow t\bar{t}$. In our numerical calculation, we will use
CTEQ6L PDF [16] for $f_{i/p} (x,\mu_{F})$ and assume the
factorization scale $\mu_{F}=m_{t}$.

The $t\bar{t}$ spin correlations manifest themselves in decay
angular correlations, which are to be measured with respect to the
chosen reference axes. If the $t(\bar{t})$ decays semileptonically
$t\rightarrow bl^+\nu_l(\bar{t}\rightarrow \bar{b}l^-\nu_l)$, the
charged lepton $l$ is the best spin analyzer [17]. A useful
observable is the following double differential distribution [4, 5,
6]:
\begin{equation}
\frac{1}{\sigma}\frac{d\sigma^{2}}{dcos\theta_{\ell^{+}}
dcos\theta_{\ell^{-}}}=\frac{1}{4}(1+B_{1}cos\theta_{\ell^{+}}
+B_{2}cos\theta_{\ell^{-}}-Ccos\theta_{\ell^{+}}
cos\theta_{\ell^{-}}).
\end{equation}
Where $\sigma$ is the cross section of the process $pp\rightarrow
t\bar{t}X\rightarrow l^+l^-X$ and $\theta_{\ell^{+}}$
$(\theta_{\ell^{-}})$ expresses the angle between the $t(\bar{t})$
spin axis and the direction of flight of the lepton $l^+(l^-)$ at
the $t(\bar{t})$ rest frame. In this paper, we choose the helicity
basis to analyze the $t\bar{t}$ spin correlations at the LHC. In
this basis, the $t(\bar{t})$ spin axis is regarded as the direction
of motion of the top (antitop) in the $t\bar{t}$ center-of-mass
system. The coefficients $B_{1}$ and $B_{2}$ are associated with a
polarization of the top and antitop quarks, and $C$ reflects the the
strength of the $t\bar{t}$ spin correlations. In this paper we focus
on investigating $C$, which can be expressed as:
\begin{equation}
C=-k_{l^+}k_{l^-}A,
\end{equation}
where $k_{l^+}$ and $k_{l^-}$ are the $t$ and $\bar{t}$
spin-analyzing powers and their values can be written as
$k_{l^+}$=$-k_{l^-}$=1 at leading order. The parameter $A$ denotes
the double spin asymmetry, which is defined as:
\begin{equation}
A=4\, \frac{\sigma(++)+\sigma(--) - \sigma(+-) - \sigma(-+)}
{\sigma(++) + \sigma(--) + \sigma(+-) +  \sigma(-+)},
\end{equation}
where $\sigma(+-)$ denotes the cross section for
$\cos\theta_{l^+}>0$ and $\cos\theta_{l^-}<0$, etc.

The total matrix element squared for the process $pp\rightarrow
t\bar{t}+X\rightarrow l^+l^-+X$ is given by:
\begin{equation}
{|M|}^2\propto
Tr[\rho^{l^+}R^i\bar{\rho}^{l^-}]=\rho^{l^+}_{\alpha'\alpha}R^i_{\alpha\beta,
\alpha'\beta'}\bar{\rho}^{l^-}_{\beta'\beta}
\end{equation}
in the narrow-width approximation for the top quark, where $\alpha'
\alpha$ and $\beta' \beta$ are the spin labels of the $t$ and
$\bar{t}$ quarks, respectively. The matrices $\rho^{l^+}$ and
$\bar{\rho}^{l^-}$ are the density matrices corresponding to the
decays $t\rightarrow l^+$ and $\bar{t}\rightarrow l^-$,
respectively. They can be written as:
\begin{eqnarray}
&& \rho^{l^+}_{\alpha'\alpha}=M(t_\alpha\rightarrow
bl^+\nu_l)M^*(t_{\alpha'}\rightarrow bl^+\nu_l),\nonumber \\
&&
\rho^{l^-}_{\beta'\beta}=M(\bar{t}_\beta\rightarrow\bar{b}l^-\bar{\nu}_l)M^*
(\bar{t}_{\beta'}\rightarrow\bar{b}l^-\bar{\nu}_l).
\end{eqnarray}
$R^i_{\alpha\beta,\alpha'\beta'}$ is the $t\bar{t}$ production
density matrice through the process $i$ in Eq.(3):
\begin{equation}
R^i_{\alpha\beta,\alpha'\beta'}=\sum_{initial~spin}M(i\rightarrow
t_\alpha\bar{t}_\beta)M^*(i\rightarrow t_{\alpha'}\bar{t}_{\beta'}),
\end{equation}
where $M(i\rightarrow t_\alpha\bar{t}_\beta)$ are the invariant
amplitudes given in Eqs.(5)-(10).

In the following section, we will use above formula to calculate
some measurable quantities related to the $t\bar{t}$ production at
the LHC.
\\\\
\noindent{\bf 3. The numerical results}
\\

In our numerical estimation, we will take $m_t=172.7GeV$,
$m_b=4.5GeV$, and $\alpha_s=0.1074$ [18]. Except for these SM input
parameters, the contributions of the scalars predicted by the TC2
model to the $t\bar{t}$ production cross section are dependent on
the free parameters $\varepsilon$, the masses of the top-pions and
top-Higgs. The free parameter $\varepsilon$ parameterizes the
portion of the extended technicolor contribution to the top quark
mass. Numerical analysis shows that, with reasonable choice of other
input parameters, $\varepsilon$ with order $10^{-2}\sim  10^{-1}$
may induce top-pions as massive as the top quark [10]. Precise value
of $\varepsilon$ may be obtained by elaborately measuring the
coupling strength between top-pion/top-Higgs and top quarks at the
next generation linear colliders. From the theoretical point of
view, $\varepsilon$ with value from 0.01 to 0.1 is favored. In this
paper, we will assume that its value is in the rang of
$0.03\sim0.1$. Since the mass splitting between neutral and charged
top-pions is very small, we assume $m_{\pi^{0}_t}=m_{\pi^{\pm}_t}$.
The top-pion mass is model-dependent and is usually of a few hundred
$GeV$ [2]. About the top-Higgs mass, Ref. [13] gives a lower bound
of about $2m_t$, but it is an approximate analysis and the mass
below $t\bar{t}$ threshold is also possible [19]. On the
experimental side, the current experiments have restricted the
masses of the charged top-pions. For example, the absence of $t
\rightarrow \pi^{+}_{t} b$  implies that $ \pi^{+}_{t}>165GeV$ [20]
and $R_{b}$ analysis yields $ m_{\pi^{+}_{t}}>220GeV$[21]. For the
masses of neutral top-pion and top-Higgs, the experimental
restrictions on them are rather weak. So, in our numerical
estimation, we will take $m_{\pi^{0}_t}=m_{\pi^{\pm}_t}=m_{h^0_t}=M$
and assume that the value of $M$ is in the range of $200GeV \sim
450GeV$.

To see whether the effects of these new scalar particles on the top
quark pair production can be detected via measuring observables at
the LHC, we define the relative correction parameters as:
\begin{eqnarray}
R_{1}&=&\mid\frac{\sigma_{tot}-\sigma_{SM}}{\sigma_{SM}}\mid,
\hspace{1.8cm}R_{2}~=~\mid\frac{A_{tot}-A_{SM}}{A_{SM}}\mid\nonumber,
\end{eqnarray}
\vspace{-1.0cm}
\begin{eqnarray}
R_{3}(P_T)&=&\mid\frac{d\sigma_{tot}/dp_{T}-d\sigma_{SM}/dp_{T}}{d\sigma_{SM}/dp_{T}}\mid
\nonumber,\\
R_{4}(M_{t\bar{t}})&=&\mid\frac{d\sigma_{tot}/dM_{t\bar{t}}-d\sigma_{SM}/dM_{t\bar{t}}}
{d\sigma_{SM}/dM_{t\bar{t}}}\mid.
\end{eqnarray}
Here the total $t\bar{t}$ production cross section $\sigma_{tot}$
includes the contributions coming from both the SM and the new
scalars predicted by the TC2 model. $P_T$ and $M_{t\bar{t}}$
represent the top quark transverse momentum and the $t\bar{t}$
invariant mass, respectively.
\begin{figure}[htb]
\vspace{-0.5cm}
\begin{center}
 \epsfig{file=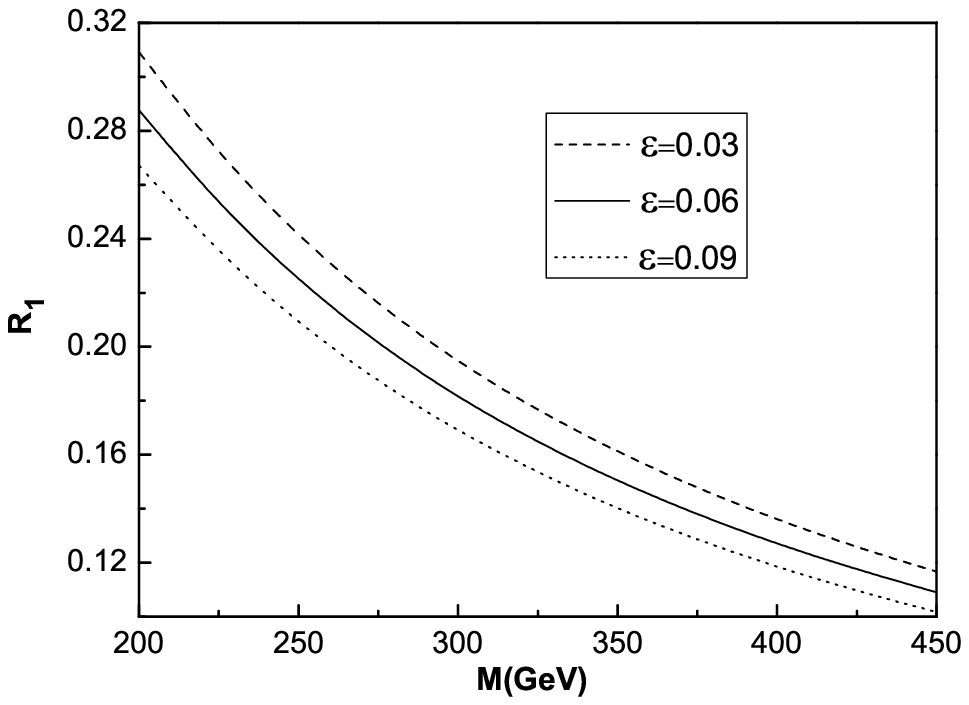,width=220pt,height=200pt}
\put(-116,-10){ (a)}\put(110,-10){ (b)}
 \hspace{0cm}\vspace{-0.25cm}
\epsfig{file=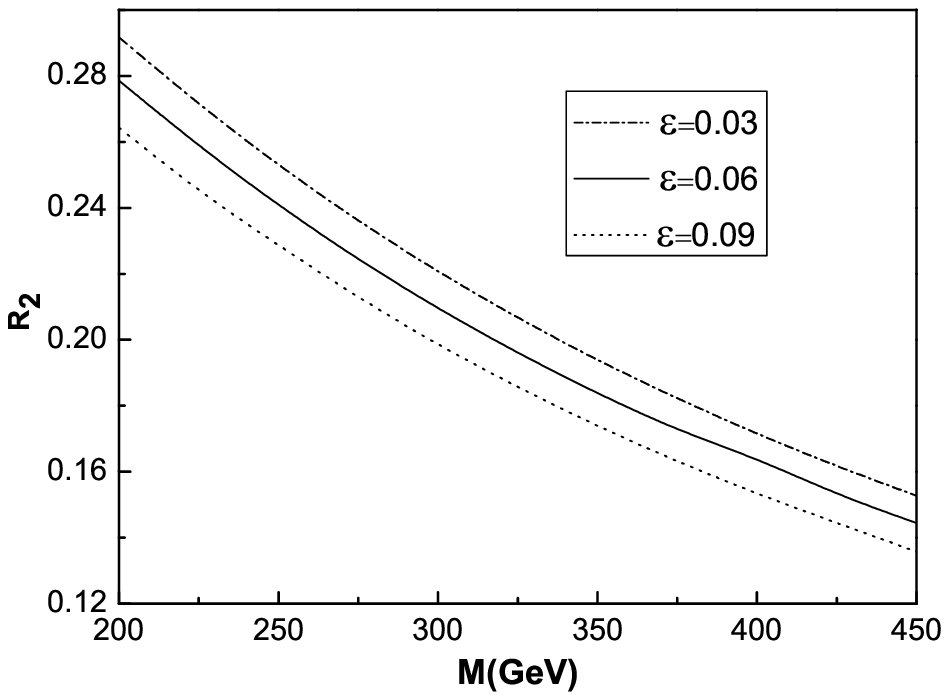,width=220pt,height=200pt} \hspace{-0.5cm}
 \hspace{10cm}\vspace{-1cm}
\vspace{0.5cm} \caption{The relative correction parameters $R_1$ and
$R_2$ are plotted as functions of the \hspace*{1.7cm} mass
parameter $M$ for three values of the parameter $\varepsilon$.}
\label{ee}
\end{center}
\end{figure}
\begin{figure}[htb]
\begin{center}
\vspace{-0.5cm}
 \epsfig{file=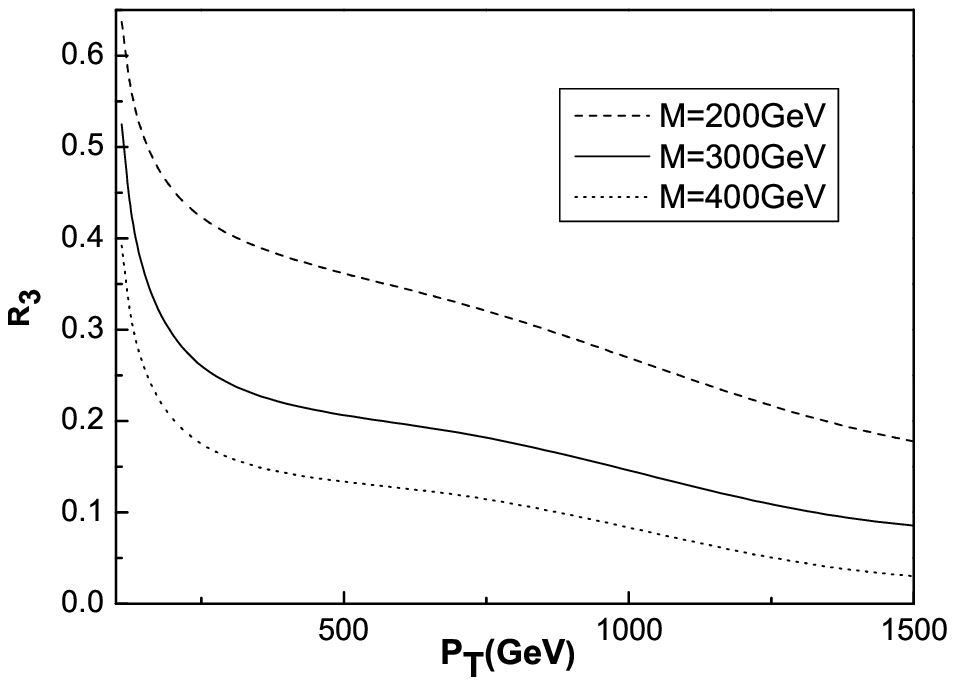,width=220pt,height=200pt}
\put(-115,-10){ (a)}\put(117,-10){ (b)}
 \hspace{0cm}\vspace{-0.25cm}
\epsfig{file=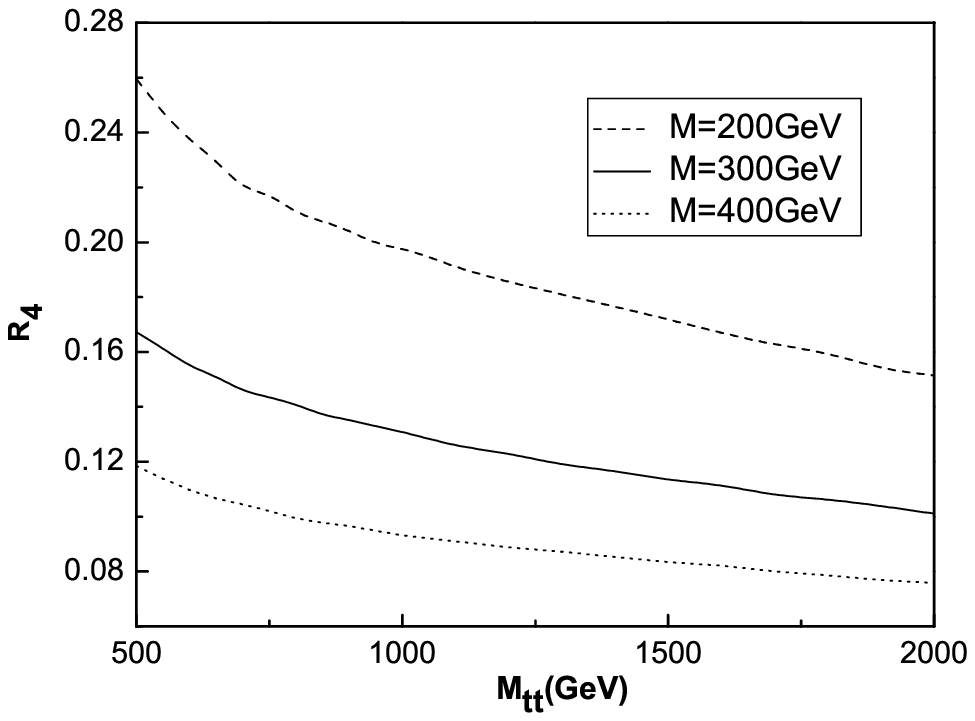,width=220pt,height=200pt} \hspace{-0.5cm}
 \hspace{10cm}\vspace{-1cm}
\vspace{0.5cm}
 \caption{The relative correction parameter $R_3(R_4)$ as a function of the transverse
 \hspace*{2.1cm}momentum $P_T$(invariant mass $M_{tt}$) for $\varepsilon=0.05$ and three values
  of the mass \hspace*{2.1cm}parameter $M$.}
 \label{ee}
\end{center}
\end{figure}

Our numerical results show that, in the case of
$m_{\pi^{0}_t}=m_{\pi^{\pm}_t}=m_{h^0_t}$, the contributions of the
neutral top-pion $\pi^{0}_t$ to the relative correction parameters
$R_{i}$ are at the same order with those for the top-Higgs $h^0_t$.
However, the contributions of the charged top-pions $\pi^{\pm}_t$
are smaller than those for $\pi^{0}_t$ or $h^0_t$ at least by one
order of magnitude. This is because, for $\pi^{\pm}_t$, the fermion
in each loop is bottom quark and furthermore there is no the
s-channel Feynman diagram Fig.1(i). Summing up all of these
contributions, we can obtain the contributions of the scalars
predicted by the TC2 model to $R_{i}$, as shown in Fig.3 and Fig.4,
in which we have taken the c.m. energy $\sqrt{s}$=$14TeV$. From
these figures one can see that these scalar particles can indeed
produce significant corrections to the observables, which are
related the top quark pair production at the LHC. The correction
effects decrease as the mass parameter $M$ and the parameter
$\varepsilon$ increasing. In wide range of the parameter space, the
correction effects of the new scalars to the $t\bar{t}$ production
cross section $\sigma(t\bar{t})$ are significant large, and the
value of the relative correction parameter $R_1$ is larger than
$10\%$, which might be detected at the LHC. For the masses of the
scalars equaling to $300GeV$ and $\varepsilon=0.05$, the values of
the relative correction parameters $R_3$ and $R_4$ are in the ranges
of $9\%\sim 52\%$ and $10\%\sim 17\%$, respectively.  For
$200GeV\leq M\leq450GeV$ and $0.03\leq \varepsilon\leq0.09$, the
values of the parameter $R_2$ is in the range of $14\%\sim29\%$.
Ref. [22] has shown that the spin asymmetry $A$ of the top-antitop
pairs in the SM will be measured with a precision of about $6\%$
after one LHC year at low luminosity ($10fb^{-1}$). Thus, the
correction effects of the scalars to the spin asymmetry $A$ should
be detected.

Certainly, the scalar particles predicted by the TC2 models can also
produce correction effects on the observables, which are related the
top quark pair production at the Tevatron. However, since the
partonic process  $q\bar{q}\rightarrow t\bar{t}$ is the dominate
process at the Tevatron, the contributions of these new particles to
the correlative observables  are smaller than those for the LHC. For
instance, in wide range of the parameter space of the TC2 model, the
relative correction value of these new scalars to the $t\bar{t}$
production cross section $\sigma(t\bar{t})$  is in the range of
$3\%\sim 11\%$ at the Tevatron.

 In conclusion, we have considered the contributions of the new scalars predicted by
the TC2 model to the $t\bar{t}$ production and the $t\bar{t}$ spin
correlations at the LHC. Our numerical results show that these new
particles can generate significant corrections to some correlative
observables. The LHC might detect these correction effects in near
future. Thus, one can use the process $pp\rightarrow t\bar{t}+X$ to
test the possible signatures of these new scalars at the LHC.
Furthermore, most of the new physics models in the topcolor scenario
predict the existence of the neutral and charged scalars, which have
similar features as those for the TC2 model. So our conclusions are
apply to the topcolor scenario.

\section*{Acknowledgments} \hspace{5mm}
J. Y. Liu thanks Zong-Guo Si for helpful discussions. This work is
supported in part by the National Natural Science Foundation of
China under Grant No.10975067, the Specialized Research Fund for the
Doctoral Program of Higher Education(SRFDP) (No.200801650002), the
Natural Science Foundation of Liaoning Science
Committee(No.2008\\2148), and the Foundation of Liaoning Educational
Committee(No.2007T086).
 \vspace{1.0cm}

\newpage
\noindent{\bf \large Appendix :}

In this Appendix, we list the form factors of the invariant
amplitudes for the partonic processes $g(p_{1}) g(p_{2})\rightarrow
t(k_{1}, s_{t})\bar{t}(k_{2}, s_{\bar{t}})$ and $q(p '_{1})
\bar{q}(p '_{2})\rightarrow t(k_{1}, s_{t})\bar{t}(k_{2},
s_{\bar{t}})$. The counterterms $\delta Z_{V}$, $\delta Z_{A}$, and
$\delta m_{t}$ coming from the top quark self-energy contributions
are listed in Appendix A6.

{\bf A1. Form factors appearing in $M_a$}
\begin{eqnarray}
A_{a}&=&1+\frac{1}{16\pi^{2}}\frac{g_{A}^{2}+g_{V}^{2}}{2m_{t}^{2}}[A_{0}(M^{2})
-A_{0}(m_{t}^{2})+(2m_{t}^{2}-M^{2})]B_{0}(m_{t}^{2},M^2,m_{t}^{2})
+\frac{1}{16\pi^{2}}\nonumber\\&&[(g_{A}^{2}+g_{V}^{2})(M^{2}-2m_{t}^{2})-
(g_{A}^{2}-g_{V}^{2})2m_{t}^2]B^{'}_{0}(m_{t}^{2})+\{-\frac{g_Vg_V^*}{16\pi^2}
\{\frac{\hat{s}}{2(\hat{s}-4m_t^2)}
\nonumber\\&&+\frac{2}{\hat{s}-4m_t^2}[A_0(m_t^2)-A_0(M^2)]+
B_0(\hat{s}^2,m_t^2,m_t^2)\frac{1}{2(\hat{s}-4m_t^2)^2}[-48m_t^4
+(-16m_t^2\nonumber\\&&+16M^2+14\hat{s})m_t^2+8m_t^2\hat{s}
-\hat{s}^2-2m_t^2\hat{s}+2M^2\hat{s}]+\frac{1}{2(\hat{s}-4m_t^2)^2}
B_0(m_t^2,M^2,m_t^2)\nonumber\\&&
[32m_t^4+(32m_t^2-32M^2-6\hat{s}^2)m_t^2-8m_t^2\hat{s}-2\hat{s}(m_t^2-M^2)]
+\frac{1}{2(\hat{s}-4m_t^2)^2}\nonumber\\&&C_0(0,m_{t}^{2},p_{1}p_{2s},m_{t}^{2},
m_{t}^{2},M^{2})
[48m_t^6+(32m_t^2-32M^2-6\hat{s})m_t^4+(32m_t^3\nonumber\\&&-32m_tM^2
-24m_t\hat{s})m_t^3+(16m_t^4+20M^2\hat{s}+16M^4-28m_t^2\hat{s}
+2\hat{s}^2-32m_t^2M^2
)m_t^2\nonumber\\&&+(4m_t\hat{s}^2-8m_t^3\hat{s}+8m_tM^2\hat{s})m_t+2m_t^2\hat{s}
+2m_t^4\hat{s}+2M^4\hat{s}-4m_t^2M^2\hat{s}]\}-\frac{g_Ag_A^*}{16\pi^2}\nonumber\\&&
\{\frac{\hat{s}}{2(\hat{s}-4m_t^2)}
+\frac{2}{\hat{s}-4m_t^2}[A_0(m_t^2)-A_0(M^2)]+\frac{1}{2(\hat{s}-4m_t^2)^2}
B_0(\hat{s}^2,m_t^2,m_t^2)\nonumber\\&&[16m_t^4
+(16M^2-16m_t^2+14\hat{s})m_t^2-8m_t^2\hat{s}
-\hat{s}^2-2m_t^2\hat{s}+2M^2\hat{s}]\nonumber\\&&\frac{1}{2(\hat{s}-4m_t^2)^2}
B_0(m_t^2,M^2,m_t^2)
[-32m_t^4+(32m_t^2-32M^2-6\hat{s}^2)m_t^2+8m_t^2\hat{s}
\nonumber\\&&-2\hat{s}(m_t^2-M^2)]
+\frac{1}{2(\hat{s}-4m_t^2)^2}C_0(0,m_{t}^{2},p_{1}p_{2s},m_{t}^{2},m_{t}^{2},M^{2})
[48m_t^6\nonumber\\&&+(32m_t^2-32M^2-6\hat{s})m_t^4+(32m_t^3+32m_tM^2
-24m_t\hat{s})m_t^3\nonumber\\&&+(16m_t^4+20M^2\hat{s}+16M^4-28m_t^2\hat{s}
+2\hat{s}^2-32m_t^2M^2
)m_t^2+(4m_t\hat{s}^2-8m_t^3\hat{s}\nonumber\\&&+8m_tM^2\hat{s})m_t+2m_t^2\hat{s}
+2m_t^4\hat{s}+2M^4\hat{s}-4m_t^2M^2\hat{s}]\}\},
\end{eqnarray}
\begin{eqnarray}
B_{a}&=&\frac{g_Vg_V^*}{16\pi^2}\{\frac{m_t}{\hat{s}-4m_t^2}
+\frac{1}{m_t(\hat{s}-4m_t^2)}[A_0(m_t^2)-A_0(M^2)]+B_0(\hat{s},
m_t^2,m_t^2)\frac{1}{(\hat{s}-4m_{t}^{2})^{2}}
\nonumber\\&&[2m_{t}^{3}-8m_{t}^{3}+(-6m_{t}^{2}+6M^{2}
+\hat{s})m_{t}+2m_{t}\hat{s}]+B_0(m_t^2,M^2,m_t^2,)
\frac{1}{m_{t}(\hat{s}-4m_{t}^{2})^{2}} \nonumber\\&&[6m_{t}^{4}
+(10m_{t}^{2}-10M^{2}-\hat{s})m_{t}^{2}-2m_{t}^2\hat{s}
+(M^{2}-m_{t}^{2})\hat{s}]\nonumber\\&&
+C_0(0,m_{t}^{2},p_{1}p_{2s},m_{t}^{2},m_{t}^{2},M^{2})
\frac{(-2)}{(\hat{s}-4m_{t}^{2})^{2}}[5m_{t}^{5}
+(2m_{t}^{2}+2M^{2}-\hat{s})m_{t}^{3}\nonumber\\&&-
(4m_{t}^{2}-4M^{2}+\hat{s})m_{t}^{3}]
+[-3m_{t}^{4}+m_{t}^{2}(6M^{2}+\hat{s})-3M^{4}-2M^{2}\hat{s}]m_{t}
\nonumber\\&&+m_{t}(m_{t}^{2}-M^{2})\hat{s}\}+\frac{g_Ag_A^*}{16\pi^2}
\{\frac{m_t}{\hat{s}-4m_t^2}
+\frac{1}{m_t(\hat{s}-4m_t^2)}[A_0(m_t^2)-A_0(M^2)]\nonumber\\&&
+B_0(\hat{s}, m_t^2,m_t^2)\frac{1}{(\hat{s}-4m_{t}^{2})^{2}}
[2m_{t}^{3}+8m_{t}^{3}+(-6m_{t}^{2}+6M^{2}
+\hat{s})m_{t}-2m_{t}\hat{s}]\nonumber\\&& +B_0(m_t^2,M^2,m_t^2,)
\frac{1}{m_{t}(\hat{s}-4m_{t}^{2})^{2}}[-10m_{t}^{4}
+(10m_{t}^{2}-10M^{2}-\hat{s})m_{t}^{2}+2m_{t}^2\hat{s}
\nonumber\\&&+(M^{2}-m_{t}^{2})\hat{s}]+
C_0(0,m_{t}^{2},p_{1}p_{2s},m_{t}^{2},m_{t}^{2},M^{2})
\frac{(-2)}{(\hat{s}-4m_{t}^{2})^{2}}[-3m_{t}^{5}
+(2m_{t}^{2}+2M^{2}\nonumber\\&&-\hat{s})m_{t}^{3}
-(4m_{t}^{2}-4M^{2}+\hat{s})m_{t}^{3}]
+[-3m_{t}^{4}+m_{t}^{2}(6M^{2}+\hat{s})-3M^{4}-2M^{2}\hat{s}]m_{t}
\nonumber\\&&-m_{t}(m_{t}^{2}-M^{2})\hat{s}\}\},
\end{eqnarray}
\begin{eqnarray}
C_{a}&=&-\frac{g_Vg_A^*}{16\pi^2}\{-1+\frac{1}{\hat{s}-4m_t^2}B_0(\hat{s},
m_t^2,m_t^2)(4m_t^2-2M^2+\hat{s})+B_0(m_t^2,M^2,m_t^2)\nonumber\\&&(\frac{-2}
{\hat{s}-4m_t^2})(4m_t^2+M^2)+C_0(0,m_{t}^{2},p_{1}p_{2s},m_{t}^{2},m_{t}^{2},M^{2})(\frac{-2}
{\hat{s}-4m_t^2})(4m_t^2+M^2)\nonumber\\&&[2m_t^4-(2m_t^2+2M^2+\hat{s})m_t^2
+M^4-2m_t^2M^2+m_t^2\hat{s}]\}.
\end{eqnarray}
Here $\hat{s}=(p_1+p_2)^2.$  $C_{0}$ is the three point scalar
function [15]. We also introduce the following shorthhand notation:
\begin{eqnarray}
B_{0}^{'}(m_{t}^{2})\equiv \frac{\partial}{\partial
P^{2}}B_{0}(m_{t}^{2};M^{2},m_{t}^{2})\mid_{p^{2}=m_{t}^{2}}.
\end{eqnarray}

 {\bf A2. Form factors appearing in $M_b$}
\begin{eqnarray}
A_{b}&=&\frac{g_{A}^{2}+g_{V}^{2}}{32\pi^{2}m_{t}^{2}}[A_{0}(M^{2})
-B_{0}(m_{t}^{2},M^2,m_{t}^{2})A_{0}(m_{t}^{2})+(2m_{t}^{2}-M^{2})]
+\frac{1}{16\pi^{2}}[(g_{A}^{2}+g_{V}^{2})\nonumber\\&&(M^{2}-2m_{t}^{2})-
(g_{A}^{2}-g_{V}^{2})2m_{t}^2]B^{'}_{0}(m_{t}^{2})+\{-\frac{g_Vg_V^*}{16\pi^2}
\{\frac{\hat{s}}{2(\hat{s}-4m_t^2)}
+\frac{2}{\hat{s}-4m_t^2}\nonumber\\&&[A_0(m_t^2)-A_0(M^2)]+\frac{1}{2(\hat{s}-4m_t^2)^2}
B_0(\hat{s}^2,m_t^2,m_t^2)[-48m_t^4
+(-16m_t^2\nonumber\\&&+16M^2+14\hat{s})m_t^2+8m_t^2\hat{s}
-\hat{s}^2-2m_t^2\hat{s}+2M^2\hat{s}]+\frac{1}{2(\hat{s}-4m_t^2)^2}
B_0(m_t^2,M^2,m_t^2)\nonumber\\&&
[32m_t^4+(32m_t^2-32M^2-6\hat{s}^2)m_t^2-8m_t^2\hat{s}-2\hat{s}(m_t^2-M^2)]
+\frac{1}{2(\hat{s}-4m_t^2)^2}\nonumber\\&&C_0(0,m_{t}^{2},p_{1}p_{2s},m_{t}^{2},
m_{t}^{2},M^{2})
[48m_t^6+(32m_t^2-32M^2-6\hat{s})m_t^4+(32m_t^3\nonumber\\&&-32m_tM^2
-24m_t\hat{s})m_t^3+(16m_t^4+20M^2\hat{s}+16M^4-28m_t^2\hat{s}
+2\hat{s}^2-32m_t^2M^2
)m_t^2 \nonumber\\
&&+(4m_t\hat{s}^2-8m_t^3\hat{s}+8m_tM^2\hat{s})m_t+2m_t^2\hat{s}
+2m_t^4\hat{s}+2M^4\hat{s}-4m_t^2M^2\hat{s}]\}-\frac{g_Ag_A^*}{16\pi^2}\nonumber
\end{eqnarray}
\begin{eqnarray}
&& \{\frac{\hat{s}}{2(\hat{s}-4m_t^2)}
+\frac{2}{\hat{s}-4m_t^2}[A_0(m_t^2)-A_0(M^2)]+\frac{1}{2(\hat{s}-4m_t^2)^2}
B_0(\hat{s}^2,m_t^2,m_t^2)\nonumber\\&&[16m_t^4
+(16M^2-16m_t^2+14\hat{s})m_t^2-8m_t^2\hat{s}
-\hat{s}^2-2m_t^2\hat{s}+2M^2\hat{s}]\frac{1}{2(\hat{s}-4m_t^2)^2}\nonumber\\&&
B_0(m_t^2,M^2,m_t^2)
[-32m_t^4+(32m_t^2-32M^2-6\hat{s}^2)m_t^2+8m_t^2\hat{s}
-2\hat{s}(m_t^2-M^2)]\nonumber\\&&
+\frac{1}{2(\hat{s}-4m_t^2)^2}C_0(0,m_{t}^{2},p_{1}p_{2s},m_{t}^{2},m_{t}^{2},M^{2})
[48m_t^6+(32m_t^2-32M^2-6\hat{s})m_t^4\nonumber\\&&+(32m_t^3+32m_tM^2
-24m_t\hat{s})m_t^3+(16m_t^4+20M^2\hat{s}+16M^4-28m_t^2\hat{s}
+2\hat{s}^2\nonumber\\&&-32m_t^2M^2
)m_t^2+(4m_t\hat{s}^2-8m_t^3\hat{s}+8m_tM^2\hat{s})m_t+2m_t^2\hat{s}
+2m_t^4\hat{s}\nonumber\\&&+2M^4\hat{s}-4m_t^2M^2\hat{s}]\}+1\},
\end{eqnarray}
\begin{eqnarray}
B_{b}&=&\frac{g_Vg_V^*}{16\pi^2}\{\frac{m_t}{\hat{s}-4m_t^2}
+\frac{1}{m_t(\hat{s}-4m_t^2)}[A_0(m_t^2)-A_0(M^2)]+\frac{1}{(\hat{s}-4m_{t}^{2})^{2}}B_0(\hat{s},
m_t^2,m_t^2) \nonumber\\&&[2m_{t}^{3}-8m_{t}^{3}+(-6m_{t}^{2}+6M^{2}
+\hat{s})m_{t}+2m_{t}\hat{s}]+B_0(m_t^2,M^2,m_t^2,)
\frac{1}{m_{t}(\hat{s}-4m_{t}^{2})^{2}} \nonumber\\&&[6m_{t}^{4}
+(10m_{t}^{2}-10M^{2}-\hat{s})m_{t}^{2}-2m_{t}^2\hat{s}
+(M^{2}-m_{t}^{2})\hat{s}]\nonumber
\end{eqnarray}
\begin{eqnarray}
&&-\frac{2}{(\hat{s}-4m_{t}^{2})^{2}}C_0(0,m_{t}^{2},p_{1}p_{2s},m_{t}^{2},m_{t}^{2},M^{2})
[5m_{t}^{5} +(2m_{t}^{2}+2M^{2}-\hat{s})m_{t}^{3}\nonumber\\&&-
(4m_{t}^{2}-4M^{2}+\hat{s})m_{t}^{3}]
+[-3m_{t}^{4}+m_{t}^{2}(6M^{2}+\hat{s})-3M^{4}-2M^{2}\hat{s}]m_{t}
\nonumber\\&&+m_{t}(m_{t}^{2}-M^{2})\hat{s}\}+\frac{g_Ag_A^*}{16\pi^2}
\{\frac{m_t}{\hat{s}-4m_t^2}
+\frac{1}{m_t(\hat{s}-4m_t^2)}[A_0(m_t^2)-A_0(M^2)]\nonumber\\&&
+B_0(\hat{s}, m_t^2,m_t^2)\frac{1}{(\hat{s}-4m_{t}^{2})^{2}}
[2m_{t}^{3}+8m_{t}^{3}+(-6m_{t}^{2}+6M^{2}
+\hat{s})m_{t}-2m_{t}\hat{s}]\nonumber\\&& +B_0(m_t^2,M^2,m_t^2,)
\frac{1}{m_{t}(\hat{s}-4m_{t}^{2})^{2}}[-10m_{t}^{4}
+(10m_{t}^{2}-10M^{2}-\hat{s})m_{t}^{2}+2m_{t}^2\hat{s}
\nonumber\\&&+(M^{2}-m_{t}^{2})\hat{s}]-
C_0(0,m_{t}^{2},p_{1}p_{2s},m_{t}^{2},m_{t}^{2},M^{2})
\frac{2}{(\hat{s}-4m_{t}^{2})^{2}}[-3m_{t}^{5}
+(2m_{t}^{2}+2M^{2}\nonumber\\&&-\hat{s})m_{t}^{3}
-(4m_{t}^{2}-4M^{2}+\hat{s})m_{t}^{3}]
+[-3m_{t}^{4}+m_{t}^{2}(6M^{2}+\hat{s})-3M^{4}-2M^{2}\hat{s}]m_{t}
\nonumber\\&&-m_{t}(m_{t}^{2}-M^{2})\hat{s}\}\},
\end{eqnarray}
\begin{eqnarray}
C_{b}&=&-\frac{g_Vg_A^*}{16\pi^2}\{-1+\frac{1}{\hat{s}-4m_t^2}(4m_t^2-2M^2+\hat{s})B_0(\hat{s},
m_t^2,m_t^2)-B_0(m_t^2,M^2,m_t^2)\nonumber\\&&(\frac{2}
{\hat{s}-4m_t^2})(4m_t^2+M^2)-\frac{2}
{\hat{s}-4m_t^2}C_0(0,m_{t}^{2},p_{1}p_{2s},m_{t}^{2},m_{t}^{2},M^{2})(4m_t^2+M^2)\nonumber\\&&[2m_t^4-(2m_t^2+2M^2+\hat{s})m_t^2
+M^4-2m_t^2M^2+m_t^2\hat{s}]\}.
\end{eqnarray}

{\bf A3. Form factors appearing in $M_c$}
\begin{eqnarray}
A_{c}&=&\frac{1}{16\pi^{2}}\frac{1}{4(p_{1}^{2}+p_{1}\cdot
p_{2})}\{(g_{A}^{2}+g_{V}^{2})[(B_{0}({0}, m_{t}^{2} ,
m_{t}^{2})-B_{0} (p_{1}p_{2s} , m_{t}^{2} ,M^{2} ))
(2m_{t}^{2}-M^{2})]\nonumber\\&&+2(p_{1}^{2}+p_{1}\cdot
p_{2})(-1+B_{0}(p_{1}p_{2s} , m_{t}^{2} ,M^{2}) -4C_{0}( 0,
m_{t}^{2},p_{1}p_{2s} , m_{t}^{2} ,m_{t}^{2},M^{2}
)m_{t}^{2})\}\nonumber\\&&-\delta Z_{V},
\end{eqnarray}
\begin{eqnarray}
B_{c}&=&-\frac{1}{16\pi^{2}}\frac{1}{4[m_{t}^{2}-2(p_{2}^{2}+p_{2}\cdot
p_{1} )](p_{2}^2+p_2\cdot p_1)^2}(g_{A}^{2}+g_{V}^{2})\{(B_{0}(0,
m_{t}^{2} , m_{t}^{2})\nonumber\\&&-B_{0}(p_{1}p_{2s} , m_{t}^{2}
,M^{2} ))(2m_{t}^{2}-M^{2}) m_{t}^{2}- 2[2B_{0}(p_{1}p_{2s} ,
m_{t}^{2} ,M^{2})m_{t}^{2}-B_{0}(m_{t}^{2},m_{t}^{2} ,
M^{2})\nonumber\\&&m_{t}^{2}-2B_{0}(p_{1}p_{2s} , m_{t}^{2} ,M^{2}
)M^{2}+2B_{0}(m_{t}^{2},m_{t}^{2} , M^{2})M^{2}+ m_{t}^{2}+
B_{0}(p_{1}p_{2s} , m_{t}^{2}
,M^{2})m_{t}^{2}\nonumber\\
&&-B_{0}(m_{t}^{2},m_{t}^{2} ,M^{2})m_{t}^{2}- 2C_{0}( 0,
m_{t}^{2},p_{1}p_{2s} ,m_{t}^{2}
,m_{t}^{2},M^{2})m_{t}^{4}+A_{0}(m_{t}^{2})+
A_{0}(M^{2})]\nonumber\\&&(p_{1}^2+p_1\cdot p_2)+ 4(1+2C_{0}( 0,
m_{t}^{2},p_{1}p_{2s} , m_{t}^{2} ,m_{t}^{2},M^{2}
)m_{t}^{2})(p_{2}^2+p_2\cdot p_1)^{2}\},
\end{eqnarray}
\begin{eqnarray}
C_{c}&=&\frac{1}{16\pi^{2}}\frac{1}{(p_{1}^2+p_1\cdot
p_2)[m_{t}^{2}-2(p_{1}^2+p_1\cdot p_2)]m_{t}}\{(B_{0}(p_{1}p_{2s} ,
m_{t}^{2} ,M^{2} )
-B_{0}(m_{t}^{2},m_{t}^{2},M^{2}))\nonumber\\&&m_{t}^{2}(2m_{t}^{2}-M^{2})g_{A}g_{V}
+2(B_{0}(m_{t}^{2},m_{t}^{2},M^{2})m_{t}^{2}-B_{0}(m_{t}^{2},m_{t}^{2}
, M^{2})M^{2}\nonumber\\&&-B_{0}(p_{1}p_{2s} , m_{t}^{2} ,M^{2}
)m_{t}^{2}+B_{0}(m_{t}^{2},m_{t}^{2} , M^{2})m_{t}^{2}
-A_{0}(m_{t}^{2})+A_{0}M^{2})\nonumber\\&&(p_{1}^2+p_1\cdot p_2))\},
\\\nonumber\\
D_{c}&= &\frac{1}{16\pi^{2}}\frac{-1}{2[m_{t}^{2}-2(p_{2}^2+p_2\cdot
p_1)](p_{2}^2+p_2\cdot p_1)^{2}}[(B_{0}(0, m_{t}^{2} ,
m_{t}^{2})-B_{0}(p_{1}p_{2s}
, m_{t}^{2} ,M^{2} ))\nonumber\\
&&m_{t}^{2}(2m_{t}^{2}-M^{2})-2(B_{0}(0, m_{t}^{2} ,
m_{t}^{2})m_{t}^{2}-2B_{0}(p_{1}p_{2s} , m_{t}^{2} ,M^{2} )
m_{t}^{2}-B_{0}(0, m_{t}^{2} ,
m_{t}^{2})M^{2}\nonumber\\&&+2B_{0}(p_{1}p_{2s} , m_{t}^{2} ,M^{2}
)M^{2}-m_{t}^{2}+B_{0}(0, m_{t}^{2} ,
m_{t}^{2})m_{t}^{2}-B_{0}(p_{1}p_{2s} , m_{t}^{2} ,M^{2}
)m_{t}^{2}\nonumber\\
&&-2C_{0}( 0, m_{t}^{2},p_{1}p_{2s} ,
          m_{t}^{2} ,m_{t}^{2},M^{2} )m_{t}^{4}
+A_{0}(m_{t}^{2})
-A_{0}(M^{2})]g_{A}g_{V}\nonumber\\&&(p_{2}^2+p_2\cdot p_1)^{2},
\\\nonumber\\
E_{c}&=&\frac{1}{16\pi^{2}}\frac{-1}{2(p_{1}^2+p_1\cdot
       p_2)}\{-2(B_{0}(p_{1}p_{2s},m_{t}^{2},M^{2})-B_{0}(m_{t}^{2},m_{t}^{2},M^{2}))(g_{A}^{2}-g_{V}^{2})m_{t}
       \nonumber\\&&+\frac{1}{m_{t}^{3}-2m_{t}(p_{1}^2+p_1\cdot
       p_2)}[(g_{A}^{2}+g_{V}^{2})((B_{0}(p_{1}p_{2s} , m_{t}^{2} ,M^{2}
       )-B_{0}(m_{t}^{2},m_{t}^{2} , M^{2}))\nonumber\\
       &&m_{t}^{2}(2m_{t}^{2}-M^{2})+2(B_{0}(m_{t}^{2},m_{t}^{2} ,M^{2})m_{t}^{2}
       -B_{0}(m_{t}^{2},m_{t}^{2} , M^{2})M^{2}-B_{0}(p_{1}p_{2s} , m_{t}^{2} ,M^{2} )
       \nonumber\\&&m_{t}^{2}
       -B_{0}(m_{t}^{2},m_{t}^{2} , M^{2})m_{t}^{2}
       -A_{0}(m_{t}^{2})+ A_{0}(M^{2}))(p_{1}^2+p_1\cdot p_2)))]\},
\end{eqnarray}
\begin{eqnarray}
F_{c}&=&\frac{1}{16\pi^{2}}\frac{-1}{2(p_{1}^2+p_1\cdot
       p_2)}\{g_{A}g_{V}[(B_{0}(p_{1}p_{2s} , m_{t}^{2} ,M^{2} )-B_{0}(m_{t}^{2}
       ,m_{t}^{2} , M^{2}))
        (2m_{t}^{2}-M^{2})\nonumber\\&&+2(-1+B_{0}(p_{1}p_{2s} , m_{t}^{2} ,M^{2} )-4C_{0}( 0,
        m_{t}^{2},p_{1}p_{2s},
        m_{t}^{2} ,m_{t}^{2},M^{2} )m_{t}^{2})(p_{1}^2+p_1\cdot p_2)]\}
        \nonumber\\&&-\delta
        Z_{A},
\end{eqnarray}
\begin{eqnarray}
G_{c}&=&\frac{1}{16\pi^{2}}\{C_{0}( 0, m_{t}^{2},p_{1}p_{2s} ,
          m_{t}^{2},m_{t}^{2},M^{2})(g_{A}^{2}-g_{V}^{2})m_{t}-\frac{1}{2(p_1^2+p_1\cdot p_2)}
        (B_0(0,m_t,m_t)\nonumber\\&&-B_0(p_1p_{2s},m_t^2,M^2))(g_A^2+g_V^2)m_t,~~~~~~
\\\nonumber\\
H_{c}&=&\frac{1}{16\pi^{2}}\frac{1}{(p_{1}^2+p_1\cdot
      p_2)}[g_{A}g_{V}(B_{0}(p_{1}p_{2s} , m_{t}^{2} ,M^{2} )-B_{0}(m_{t}^{2},m_{T}^{2} ,
      m^{2}))m_{t}].~~~~~~~~~~~~~~~~
\end{eqnarray}
Where $p_{1}p_{2s}=m_{t}-2(p_{1}^2+p_1\cdot p_2)$ for the off-shell
$\bar{t}$ quark and $p_{1}p_{2s}=m_{t}-2(p_{2}^2+p_2\cdot p_1)$ for
 the off-shell $t$ quark.
\noindent%{\bf 2. $t$ momentum is off-shell}

{\bf A4. Form factors appearing in $M_e$}

\begin{eqnarray}
A_{e}&=&\frac{1}{16\pi^{2}}\frac{-1}{4(p_{2}^2+p_2\cdot
      p_1)}\{(g_{A}^{2}+g_{V}^{2})[(B_{0}(m_{t}^{2} ,M^{2}, m_{t}^{2})
      -B_{0}( p_{1}p_{2s}, M^{2} , m_{t}^{2}))(2m_{t}^{2}-M^{2})\nonumber\\
      &&+(2-2B_{0}( p_{1}p_{2s},M^{2} , m_{t}^{2})+4C_{0}(m_{t}^{2}+M^{2})
      (m_{t}^{2} ,0, p_{1}p_{2s} , M^{2} ,m_{t}^{2} ,m_{t}^{2}))\nonumber\\&&\frac{1}{(p_{2}^2+p_2\cdot
      p_1)}]\}
     -\delta Z_{V},
\end{eqnarray}
\begin{eqnarray}
B_{e}&=&\frac{1}{16\pi^{2}}\frac{1}{4[m_{t}^{2}-2(p_{2}^2+p_2\cdot
p_1)](p_{2}^2+p_2\cdot
      p_1)^{2}}\{[(B_{0}(m_{t}^{2} ,M^{2}, m_{t}^{2})
      -B_{0}( p_{1}p_{2s}, M^{2} , m_{t}^{2}))\nonumber\\&&(2m_{t}^{2}-M^{2}
      )m_{t}^{2}-2(B_{0}(m_{t}^{2} ,M^{2}, m_{t}^{2})m_{t}^{2}-
      B_{0}( p_{1}p_{2s},M^{2} , m_{t}^{2})m_{t}^{2}-B_{0}^{12}M^{2} \nonumber\\&&-B_{0}
      ( p_{1}p_{2s}, M^{2} , m_{t}^{2})M^{2}
     -m_{t}^{2}+
      B_{0}(m_{t}^{2} ,M^{2}, m_{t}^{2})m_{t}^{2}-
      B_{0}( p_{1}p_{2s}, M^{2} , m_{t}^{2})m_{t}^{2}\nonumber\\
      &&-2C_{0}(m_{t}^{2} ,0, p_{1}p_{2s} ,M^{2} ,
      m_{t}^{2} ,m_{t}^{2})m_{t}^{4}-A_{0}(m_{t}^{2})
      -A_{0}(M^{2}))
      (p_{2}^2+p_2\cdot p_1)\nonumber\\&&-4(1+2C_{0}
      (m_{t}^{2},0,p_{1}p_{2s},M^{2},m_{t}^{2},m_{t}^{2})m_{t}^{2})
      (p_{2}^2+p_2\cdot p_1)^{2}](g_{A}^{2}+g_{V}^{2})\},
\end{eqnarray}
\begin{eqnarray}
C_{e}&=&\frac{1}{16\pi^{2}}\frac{-1}{m_{t}[m_{t}^{2}-2(p_{2}^2+p_2\cdot
p_1)](p_{2}^2+p_2\cdot p_1)}\{g_{A}g_{V}[(B_{0}(m_{t}^{2} ,M^{2},
m_{t}^{2})\nonumber\\&&
     -B_{0}( p_{1}p_{2s}, M^{2} , m_{t}^{2}))(2m_{t}^{2}-M^{2}
     )m_{t}^{2}-2(B_{0}(m_{t}^{2} ,M^{2}, m_{t}^{2})m_{t}^{2}
     -B_{0}(m_{t}^{2} ,M^{2}, m_{t}^{2})M^{2}\nonumber\\&&+A_{0}(M^{2})
     +B_{0}(m_{t}^{2} ,M^{2}, m_{t}^{2})m_{t}^{2}-B_{0}
     ( p_{1}p_{2s}, M^{2} , m_{t}^{2})m_{t}^{2}-A_{0}
     (m_{t}^{2}))\nonumber\\&&(p_{2}^2+p_2\cdot
     p_1)]\},
\end{eqnarray}
\begin{eqnarray}
D_{e}&=&\frac{1}{16\pi^{2}}\frac{-1}{2[m_{t}^{2}-2(p_{2}^2+p_2\cdot
p_1)](p_{2}^2+p_2\cdot p_1)^{2}}\{[(B_{0}(m_{t}^{2} ,M^{2},
m_{t}^{2})-B_{0}( p_{1}p_{2s}, M^{2} , m_{t}^{2}))\nonumber\\
&&(2m_{t}^{2}-M^{2})m_{t}^{2}-2(B_{0}(m_{t}^{2} ,M^{2},
m_{t}^{2})m_{t}^{2}-B_{0}( p_{1}p_{2s},M^{2} , m_{t}^{2})m_{t}^{2}
\nonumber\\&&+B_{0}(m_{t}^{2} ,M^{2}, m_{t}^{2})M^{2}+B_{0}
      ( p_{1}p_{2s},M^{2} , m_{t}^{2})M^{2}-m_{t}^{2}
      +B_{0}(m_{t}^{2} ,M^{2}, m_{t}^{2})m_{t}^{2}
      \nonumber\\&&-B_{0}( p_{1}p_{2s},M^{2} , m_{t}^{2})m_{t}^{2}-2C_{0}
      (m_{t}^{2} ,0, p_{1}p_{2s} , M^{2} ,
          m_{t}^{2} ,m_{t}^{2})m_{t}^{4}
      +A_{0}(m_{t}^{2})-A_{0}M^{2}))\nonumber\\&&
      (p_{2}^2+p_2\cdot p_1)-4(1+2C_{0}(m_{t}^{2} ,0, p_{1}p_{2s} , M^{2} ,
          m_{t}^{2} ,m_{t}^{2})m_{t}^{2})
     (p_{2}^2+p_2\cdot p_1)^{2}]g_{A}g_{V}\},
\end{eqnarray}
\begin{eqnarray}
E_{e}&=&\frac{1}{8\pi^{2}}C_{0}(m_{t}^{2},0,p_{1}p_{2s},M^{2},m_{t}^{2},m_{t}^{2})
2g_{V}^{2}m_{t}+[(-g_{A}^{2}m_{t}-3g_{V}^{2}m_{t}) (B_{0}(m_{t}^{2}
,M^{2}, m_{t}^{2})-B_{0}\nonumber\\&&( p_{1}p_{2s}, M^{2} ,
m_{t}^{2})-2C_{0}(m_{t}^{2} ,0, p_{1}p_{2s} , M^{2} ,
          m_{t}^{2} ,m_{t}^{2})(p_{2}^2+p_2\cdot p_1))]\nonumber\\&&
          +\frac{1}{16\pi^{2}}\frac{1}{(p_{2}^2+p_2\cdot
      p_1)}\{(g_{A}^{2}+g_{V}^{2})m_{t}
      [B_{0}(m_{t}^{2} ,M^{2}, m_{t}^{2})
      -B_{0}( p_{1}p_{2s}, M^{2} , m_{t}^{2})\nonumber\\&&+
      \frac{B_{0}(m_{t}^{2} ,M^{2}, m_{t}^{2})
      (-M^{2})+A_{0}(m_{t}^{2})
      +A_{0}(M^{2})}{2m_{t}^{2}}
     \nonumber\\&&+2C_{0}(m_{t}^{2} ,0, p_{1}p_{2s} , M^{2} ,
      m_{t}^{2} ,m_{t}^{2})(p_{2}^2+p_2\cdot
      p_1)
      \nonumber\\&&+\frac{-A_{0}(m_{t}^{2})+A_{0}(M^{2})
      +B_{0}( p_{1}p_{2s}, M^{2} , m_{t}^{2})(-M^{2}
      +2(p_{2}^2+p_2\cdot p_1))}{2[m_{t}^{2}-2(p_{2}^2+p_2\cdot p_1)]}]\},
\end{eqnarray}
\begin{eqnarray}
F_{e}&=&\frac{1}{32\pi^{2}(p_{2}^2+p_2\cdot
      p_1)}\{g_{A}g_{V}[(B_{0}(m_{t}^{2} ,M^{2}, m_{t}^{2})
      -B_{0}( p_{1}p_{2s}, M^{2} , m_{t}^{2}))
(2m_{t}^{2}-M^{2})+4C_{0}\nonumber\\&&(m_{t}^{2} ,0,
p_{1}p_{2s},M^{2} , m_{t}^{2}
,m_{t}^{2})(m_{t}^{2}+M^{2})(p_{2}^2+p_2\cdot p_1)+(2-2B_{0}(
p_{1}p_{2s}, M^{2} , m_{t}^{2})]\}\nonumber\\&&-\delta Z_{A},
\\\nonumber\\
G_{e}&=&\frac{m_t}{32\pi^{2}(p_{2}^2+p_2\cdot p_1)}(B_{0}(m_{t}^{2}
,M^{2}, m_{t}^{2})-B_{0}( p_{1}p_{2s}, M^{2} , m_{t}^{2}))
      (g_{A}^{2} +g_{V}^{2})
      +(p_{2}^2+p_2\cdot p_1)\nonumber\\&&
      2C_{0}(m_{t}^{2} ,0, p_{1}p_{2s} ,M^{2} ,
          m_{t}^{2} ,m_{t}^{2})m_{t}(g_{A}^{2}-g_{V}^{2}),
\end{eqnarray}
\begin{eqnarray}
H_{e}&=&\frac{1}{16\pi^{2}}\frac{1}{(p_{2}^2+p_2\cdot
      p_1)^{2}}g_{A}g_{V}m_{t}[2(B_{0}(m_{t}^{2} ,M^{2}, m_{t}^{2})
      -B_{0}( p_{1}p_{2s}, M^{2} , m_{t}^{2}))m_{t}^{2}
      \nonumber\\&&+(B_{0}(m_{t}^{2} ,M^{2}, m_{t}^{2})-B_{0}(0, m_{t}^{2} , m_{t}^{2})
                      -2C_{0}(m_{t}^{2} ,0, p_{1}p_{2s} , M^{2} ,
          m_{t}^{2} ,m_{t}^{2})M^2)M^{2}(p_2\cdot p_1\nonumber\\&&+p_{2}^2)+2C_{0}(m_{t}^{2} ,0, p_{1}p_{2s} , M^{2} ,
          m_{t}^{2} ,m_{t}^{2})+B_{0}( p_{1}p_{2s},M^{2} , m_{t}^{2})].
\end{eqnarray}

\noindent{\bf A5. Form factors appearing in $M_i$ }
\begin{eqnarray}
A_{i}&=&4g_Vm_t\{[m_t^2-(p_1+p_2)^2+(p_1^2+p_1\cdot p_2)]C_0(p_1^2,
(p_2-p_1)^2, 0, m_t^2,
m_t^2)\nonumber\\&&[(p_1-3p_2)(p_2-p_1)]C_{12}(p_1^2, (p_2-p_1)^2,
0, m_t^2, m_t^2)-(3p_1^2+p_1\cdot p_2)\nonumber\\&&C_{11}(p_1^2,
(p_2-p_1)^2, 0, m_t^2, m_t^2)+2(p_1\cdot p_2-p_1^2)C_{11}(p_1^2,
(p_2-p_1)^2, 0, m_t^2, m_t^2)\nonumber\\&&C_{12}(p_1^2, (p_2-p_1)^2,
0, m_t^2, m_t^2)-p_1^2C_{11}^2(p_1^2, (p_2-p_1)^2, 0, m_t^2, m_t^2),
\end{eqnarray}
\begin{eqnarray}
B_{i}&=&16m_tg_VC_{22}(p_1^2, (p_2-p_1)^2, 0, m_t^2, m_t^2),
\\
C_{i}&=&8m_tg_V[2C_{23}(p_1^2, (p_2-p_1)^2, 0, m_t^2,
m_t^2)-C_{12}(p_1^2, (p_2-p_1)^2, 0, m_t^2, m_t^2)],
\\D_{i}&=&8m_tg_VC_{12}(p_1^2, (p_2-p_1)^2, 0, m_t^2, m_t^2),
\\
E_{i}&=&16m_tg_VC_{12}(p_1^2, (p_2-p_1)^2, 0, m_t^2, m_t^2),
\\
F_{i}&=&-16m_tg_VC_{11}(p_1^2, (p_2-p_1)^2, 0, m_t^2, m_t^2),
\\
G_{i}&=&8m_tg_VC_{0}(p_1^2, (p_2-p_1)^2, 0, m_t^2, m_t^2),
\\
H_{i}&=&-4im_tg_AC_{0}(p_1^2, (p_2-p_1)^2, 0, m_t^2, m_t^2),
\\
I_{i}&=&-4m_tg_VC_{0}(p_1^2, (p_2-p_1)^2, 0, m_t^2, m_t^2),
\\
J_{i}&=&4im_tg_AC_{0}(p_1^2, (p_2-p_1)^2, 0, m_t^2, m_t^2).
\end{eqnarray}

\noindent{\bf A6. The counterterms $\delta Z_{V}$, $\delta Z_{A}$,
and $\delta m_{t}$ }
\begin{eqnarray}
\delta
Z_{A}&=&\frac{1}{16\pi^{2}}\frac{g_{A}g_{V}}{m_{t}^{2}}[A_{0}(M^{2})
-A_{0}(m_{t}^{2})+(2m_{t}^{2}-M^{2})B_{0}(m_t^2,M^2,m_t^2)],
\end{eqnarray}
\begin{eqnarray}
\delta
Z_{V}&=&\frac{1}{16\pi^{2}}\frac{g_{A}^{2}+g_{V}^{2}}{2m_{t}^{2}}[A_{0}(M^{2})
-A_{0}(m_{t}^{2})+(2m_{t}^{2}-M^{2})]\nonumber\\&&B_{0}(m_{t}^{2})(m_t^2,M^2,m_t^2)
+\frac{1}{16\pi^{2}}[(g_{A}^{2}+g_{V}^{2})(M^{2}-2m_{t}^{2})\nonumber\\&&-
(g_{A}^{2}-g_{V}^{2})2m_{t}^2]B^{'}_{0}(m_{t}^{2}),
\\\nonumber\\
\delta
m_{t}&=&-\frac{1}{16\pi^{2}}[(4m_{t}^2-M^{2})(g_{A}^{2}+g_{V}^{2})
]B^{'}_{0}(m_{t}^{2})
+\frac{1}{16\pi^{2}m_{t}^{2}}\nonumber\\&&\{(g_{A}^{2}+g_{V}^{2})
[A_{0}(M)^{2}-A_{0}(m_{t}^{2})]+[(g_{A}^{2}+g_{V}^{2})\nonumber\\&&(m_{t}^{2}-M^{2})
 +(g_{A}^{2}-g_{V}^{2})m_{t}^2]B_{0}(m_t^2,M^2,m_t^2)\}.
\end{eqnarray}

\end{document}